\documentclass[showpacs,preprintnumbers,amsmath,amssymb]{revtex4-2}

\usepackage[utf8]{inputenc}
\usepackage[pdftex]{graphicx}
\usepackage{caption}
\usepackage{subcaption}
\usepackage{mathrsfs}
\usepackage[colorlinks, breaklinks, urlcolor={blue}, linkcolor={red}, citecolor={blue}]{hyperref}
\usepackage{array}
\usepackage{amsmath}
\usepackage{type1cm}
\usepackage{lettrine}
\usepackage[english]{babel}
\usepackage{lmodern}
\usepackage{microtype}
\usepackage{booktabs}
\usepackage[T1]{fontenc}
\usepackage[boxed, vlined]{algorithm2e}
\usepackage{caption}
\usepackage{braket}
\usepackage{xcolor}
\usepackage{orcidlink}
\usepackage{bm}
\usepackage{bbold}
\usepackage{upgreek}

\begin{document}

\title{Enhanced quantum synchronization of a driven qubit under non-Markovian dynamics}

\author{Po-Wen Chen}
\email{powen@nari.org.tw}
\affiliation{Department of Physics, National Atomic Research Institute, Taoyuan City 325207, Taiwan.}

\author{Chandrashekar Radhakrishnan \orcidlink{0000-0001-9721-1741}}
\email{chandrashekar.radhakrishnan@nyu.edu}
\affiliation{Department of Computer Science and Engineering, New York University Shanghai, 567
West Yangsi Road, Pudong, Shanghai 200124, China}

\author{Md.~Manirul Ali \orcidlink{0000-0002-5076-7619}}
\email{manirul@citchennai.net}
\affiliation{Centre for Quantum Science and Technology, Chennai Institute of Technology, Chennai 600069, India}

\begin{abstract}
Synchronizing a few-level quantum system is of fundamental importance to the understanding of synchronization in the deep quantum regime. We investigate quantum phase synchronization of a two-level system (qubit) driven by a semiclassical laser field, in the presence of a general non-Markovian dissipative environment. The phase preference of the qubit is demonstrated through Husimi Q-function, and the existence of a limit cycle is also shown in our system. Synchronization of the qubit is quantified using the shifted phase distribution. The signature of quantum phase synchronization {\it viz} the Arnold tongue is obtained from the maximal value of the shifted phase distribution. Two distinct types of qubit dynamics is considered depending on the reservoir correlation time being very short and a situation when bath correlation time is finite. In the Markov regime of the environment, the phase preference of the qubit goes away in the long time limit, whereas the long-time phase localization persists in the non-Markovian regime. We also plot the maximum of the shifted phase distribution in two ways: (a) by varying the detuning and laser driving strength, and (b) by varying the system-bath coupling and laser driving strength. Various system-environment parameters determine the synchronization regions and the qubit phase synchronization is shown to be enhanced in the non-Markov regime.
\end{abstract}

\maketitle

\section{Introduction}\label{sec:Introduction}

Synchronization is a natural phenomenon that occurs in a variety of physical, chemical, and biological systems and has been
extensively studied and observed in nature for many years \cite{pikovsky2001synchronization,strogatz2018nonlinear,arenas2008synchronization,van1926lxxxviii,manrubia2004emergence,osipov2007synchronization}.
For example, if an autonomous oscillating system is coupled to another such system or an external driving force, it can synchronize
its frequency and phase to the external system/driving \cite{rosenblum2003synchronization}. A well-known example of classical synchronization is the van der Pol oscillator model \cite{van1926lxxxviii,manrubia2004emergence,osipov2007synchronization}.
In recent past, van der Pol oscillator model was reformulated in terms of a quantum system \cite{lee2013quantum,walter2014quantum,lorch2016genuine,arosh2021quantum}, and it was shown that when the system is far
from the ground state, synchronization in quantum systems is analogous to classical synchronization under the influence of
noise \cite{lee2013quantum}. When we are close to the ground state, this correspondence is changed because the discreteness
of the energy levels becomes important. It is therefore interesting to study synchronization in quantum systems with a small
number of energy levels. Studying synchronization of finite dimensional quantum systems have gained momentum due
to its potential application in the field of quantum computation and quantum information. Recently, quantum synchronization
in low-dimensional systems has been investigated \cite{zhirov2008synchronization,roulet2018synchronizing,koppenhofer2019optimal,parra2020synchronization,xiao2023classical}.
Generally synchronization can be classified into forced synchronization and spontaneous synchronization
(or mutual synchronization). In the case of spontaneous synchronization \cite{orth2010dynamics,giorgi2012quantum,giorgi2013spontaneous,ameri2015mutual,hush2015spin,roulet2018quantum,
henriet2019environment,karpat2019quantum,karpat2020quantum} the interested system becomes
synchronized in the transient evolution of dynamical systems due to the interaction between the subsystems or an
external environment. In contrast to spontaneous synchronization, forced synchronization or entrainment usually emerges with
externally driven forces \cite{walter2014quantum,goychuk2006quantum,zhirov2008synchronization,roulet2018synchronizing,koppenhofer2019optimal,
parra2020synchronization,xiao2023classical}. Here we investigate the quantum phase synchronization of a two-level system in
presence of an external driving field. Synchronizing a few-level quantum system is of fundamental importance to the understanding
of synchronization in the deep quantum regime. Initially in Refs.
\cite{goychuk2006quantum,zhirov2008synchronization,giorgi2013spontaneous,ameri2015mutual}, qubits were suggested to be
the smallest possible system that can be synchronized. Later, spin-1 systems were theoretically shown to be synchronizable
\cite{roulet2018synchronizing,koppenhofer2019optimal,roulet2018quantum}, experimental demonstrations of which were
carried out subsequently \cite{laskar2020observation,koppenhofer2020quantum}. However in Ref. \cite{roulet2018synchronizing},
it was claimed that quantum synchronization is not applicable for a single qubit (spin-1/2 system) due to the lack of limit cycle.
Subsequently it was shown that limit cycle of a single qubit can be obtained and synchronization of a qubit to an external signal
is possible \cite{parra2020synchronization,xiao2023classical}. Following that, synchronization of a single qubit to an
external driving signal is experimentally demonstrated using trapped-ion system \cite{zhang2023quantum}.

In the present work, we investigate transient phase synchronization for a two-level system (qubit) driven by a
semiclassical laser field and simultaneously coupled to a dissipative environment, existence of limit cycle is
shown in our system. We discuss the phase synchronization in which the qubit synchronizes under the driving in
presence of a Markov or non-Markov environment. Our synchronization analysis for the two-level system is applicable
to a general non-Markovian environment. Recently, non-Markovian environments have drawn particular attention
in quantum science and technology when environment's correlation time is not too small compared
with the system's relaxation time in many physical systems \cite{breuer2016colloquium,de2017dynamics}.
It is interesting to investigate the relationship between non-Markovianity and quantum synchronization
\cite{eneriz2019degree,karpat2021synchronization,huang2023classical,xiao2023classical}. Here we
show that non-Markovianity has a significantly positive impact on the emergence of synchronization.
The rest of the paper is organized as follows. In Sec.~\ref{sec:Model}, we consider a widely studied
two-level system (qubit) simultaneously interacting with a dissipative
environment while being driven by an external field. The time evolution of the
reduced density matrix of the qubit is considered to be two distinct types of dynamics depending on $\left( 1\right)$ the
reservoir correlation time is being very small compared to the relaxation time of the qubit
and the dynamics corresponds to Markovian dynamics. On the other hand, $\left(2\right)$ the reservoir correlation time
is of the same order as the system relaxation time and connected with non-Markovian effects
\cite{chen2014investigating,chen2016enhanced,ali2017probing,radhakrishnan2019time,ali2024detecting}.
In Sec.~\ref{sec:Husimi}, we demonstrate the transient dynamics of the Husimi $Q$-representation in order to visualize and
characterize phase synchronization behavior in both the Markov and non-Markov regimes. In Sec.~\ref{sec:arnold}, we
consider a measure of synchronization, called shifted phase distribution, and show its dynamics
as a function of phase and detuning in the Markov and non-Markovian regimes. We also plot the maximum value
of the shifted phase distribution in two different ways: (a) by varying the detuning and laser drive strength,
and (b) by varying the system-bath coupling and laser drive strength. Signature of quantum phase synchronization
{\it viz} the Arnold tongue is demonstrated through the maximal value of the shifted phase distribution. Finally,
we present our conclusions in Sec.~\ref{sec:conclusion}.

\section{Model of laser-driven qubit and time evolution}\label{sec:Model}
A single two level system (TLS) driven by a semiclassical laser field is coupled to a dissipative environment.  In this study
we work with the rotating wave approximation (RWA) and the total Hamiltonian of the system plus environment and the external driving
is given by:
\begin{equation}
H= H_{S} + H_{R} + H_{SR} + H_{d} =
\frac{\hbar}{2} \omega _{0}{\sigma}_{z}
+\sum_{k}\hbar \omega _{k}b_{k}^{\dagger}b_{k}
+\sum_{k}\hbar \left(g_{k}{\sigma}_{+}b_{k}
+g_{k}^{\ast} {\sigma}_{-}b_{k}^{\dagger }\right)
+ i\hbar \frac{\epsilon}{4}
\left( e^{i\omega_{L}t}{\sigma}_{-} + e^{-i\omega _{L}t} {\sigma}_{+}\right),
\label{Equation1}
\end{equation}
where the ground state to excited state transition frequency is $\omega_{0}$. The strength of the semi-classical laser field is
$\epsilon$ and its driving frequency is $\omega_{L}$. The system Hamiltonian $H_{S}$ can be written in terms of the spin-raising
and spin-lowering operators $\sigma_{+} = |1 \rangle \langle 0 |$ and $\sigma_{-} = |0 \rangle \langle 1 |$.  The Hamiltonian of
the dissipative environment $H_{R}$ is described by a collection of infinite bosonic modes with the bosonic creation and annihilation
operators denoted by $b_{k}^{\dagger}$ and $b_{k}$ respectively.  The factor $g_{k}$ is the coupling strength between the system
and the $k^{th}$ mode of the environment with frequency $\omega_{k}$.  Using the unitary transformation,
$U(t)=e^{\frac{i}{2} {\sigma}_{z} \omega_{L} t}$, we transform the Hamiltonian to a frame rotating at the laser driving frequncy.
The total Hamiltonian after this transformation reads:
\begin{eqnarray}
H = H_{\text{TLS}}+H_{\text{SR}}+H_{\text{R}}
= \frac{\hbar }{2}\Delta {\sigma}_{z}+\frac{\hbar }{2}\epsilon {\sigma}_{x}
+\sum_{k}\hbar \left( g_{k} {\sigma}_{+}b_{k}e^{i\omega_{L}t}+g_{k}^{\ast} {\sigma}_{-}
b_{k}^{\dagger }e^{-i\omega_{L}t}\right) +\sum_{k}\hbar \omega _{k}b_{k}^{\dagger }b_{k},
\label{Equation2}
\end{eqnarray}
where $\Delta =\omega _{0}-\omega _{L}$ represents the detuning with the laser driving.
The two-level system Hamiltonian can precisely be diagonalized as
\begin{equation}
H_{\text{TLS}}=\frac{\hbar }{2}\Delta {\sigma}_{z}+\frac{\hbar }{2}
\epsilon {\sigma}_{x}=\frac{\hbar }{2}\delta {\bar \sigma}_{z},
\label{Equation3}
\end{equation}
where we use the basis transformation $\vert {\bar 1} \rangle = \cos \frac{\theta}{2}\left\vert 1 \right\rangle +\sin \frac{\theta }{2}
\left\vert 0 \right\rangle$, and $\vert {\bar 0} \rangle =-\sin \frac{\theta }{2} \left\vert 1 \right\rangle +\cos \frac{\theta }{2}
\left\vert 0 \right\rangle$, and we choose $\tan $ $\theta =\frac{\epsilon }{\Delta }$ and $\delta =\sqrt{\Delta ^{2}+\epsilon ^{2}}$.
In the new basis the spin operators are defined as ${\bar \sigma}_{+}=\left\vert {\bar 1} \right\rangle \left\langle {\bar 0} \right\vert$,
${\bar \sigma}_{-}=\left\vert {\bar 0} \right\rangle \left\langle {\bar 1} \right\vert$, and ${\bar \sigma}_{z}=\left\vert {\bar 1}
\right\rangle \left\langle {\bar 1} \right\vert - \left\vert {\bar 0} \right\rangle \left\langle {\bar 0} \right\vert$.
The system-reservoir interaction Hamiltonian in the interaction picture reads:
\begin{equation}
H_{\text{SR}}(t) =\sum_{k}\hbar g_{k}\left[ P_{\text{0}} {\bar \sigma}_{z}
+P_{+} {\bar \sigma}_{+} + P_{-} {\bar \sigma}_{-} \right] b_{k}~
e^{i(\omega_{L}- \omega_k)t} + \text{H.c.},
\label{Equation4}
\end{equation}
where $P_{\text{0}}=\frac{\epsilon }{2\delta },$ $P_{+}=\frac{\Delta +\delta}{2\delta },$ and
$P_{-}=\frac{\Delta -\delta }{2\delta }$. The interaction Hamiltonian can finally be expressed as
\begin{equation}
H_{\text{SR}}\left(t\right) =S(t)B\left( t\right) + \text{H.c.},
\label{Equation5}
\end{equation}
where the reservoir operator $B\left( t\right)$ and the factor $S(t)$ are defined as
\begin{equation}
B\left( t\right) =\sum_{k}g_{k}b_{k}e^{-i\omega _{k}t};  \qquad \qquad
S(t)=\left(P_{\text{0}} {\bar \sigma}_{z}+P_{+} {\bar \sigma}_{+}e^{i\delta t}+P_{-} {\bar \sigma}_{-}e^{-i\delta t}\right) e^{i\omega_{L}t}.
\label{sbparameters}
\end{equation}
The constant $P_{\text{0}}$ denotes the elastic tunnelling through the bath, whereas the $P_{+}$ and $ P_{-} $ denotes the inelastic excitation
and relaxation through the bath respectively.  For simplicity, we consider factorized initial system-environment state
$\rho_{T}\left( 0\right)=\rho \left( 0\right) \otimes \rho_{E}\left(0\right)$, and $\rho_{E}\left( 0\right)$ assumes the initial thermal
equilibrium state,
\begin{equation}
\rho _{E}\left( 0\right) =\exp \left( -\beta H_{\text{R}} \right) / {\rm Tr}
\left[ \exp \left( -\beta H_{\text{R}} \right) \right].
\label{Equation8}
\end{equation}
Here $\beta=1/k_{B}T$ with $k_{B}$ is the Boltzmann constant and $T$ is the temperature. The second-order perturbative master equation in the
Schr\"{o}dinger picture for the two-level system is given by
\begin{eqnarray}
\frac{d\rho}{dt} &=&-\frac{i\delta }{2}\left[ {\sigma}_{z},\rho \right]  \notag \\
&& + \Big[~\Gamma_{1}(t) \left\{ P_{\text{0}}^{2}\left( {\sigma}_{z}\rho
{\sigma}_{z} - {\sigma}_{z} {\sigma}_{z}\rho \right) + P_{\text{0}}
P_{+}\left({\sigma}_{z}\rho {\sigma}_{+} - {\sigma}_{+} {\sigma}_{z}\rho \right)
+ P_{\text{0}}P_{-}\left({\sigma}_{z} \rho {\sigma}_{-} - {\sigma}_{-}
{\sigma}_{z}\rho \right) \right\} + \notag \\
&& ~~~~~\Gamma_{2}(t) \left\{ P_{-}P_{\text{0}}\left( {\sigma}
_{+}\rho {\sigma}_{z} - {\sigma}_{z} {\sigma}_{+}\rho \right)
+P_{-}P_{+}\left( {\sigma}_{+} \rho {\sigma}_{+} - {\sigma}_{+}
{\sigma}_{+}\rho \right) + P_{-}^{2}\left( {\sigma}_{+} \rho {\sigma}
_{-} - {\sigma}_{-} {\sigma}_{+} \rho \right) \right\} + \notag \\
&& ~~~~~\Gamma_{3}(t) \left\{ P_{+} P_{\text{0}}
\left( {\sigma}_{-} \rho {\sigma}_{z} - {\sigma}_{z} {\sigma}_{-}\rho \right)
+ P_{+}^{2}\left( {\sigma}_{-}\rho {\sigma}_{+} - {\sigma}_{+}
{\sigma}_{-}\rho \right) +P_{+}P_{-}\left( {\sigma}_{-}\rho {\sigma}
_{-} - {\sigma}_{-} {\sigma}_{-}\rho \right) \right\} + \rm{H.c.} \Big] ,
\label{Equation9}
\end{eqnarray}
where we have omitted the bars from the spin operators and it is implicit that the Pauli spin operators
are now in the new basis. The time dependent coefficients in the master equation are denoted by
\begin{eqnarray}
\Gamma_{1}(t) &=&\int_{0}^{t}d\tau \int_{0}^{\infty }d\omega
J\left( \omega \right) e^{-i\left( \omega -\omega _{L}\right) \left( t-\tau\right) },
\label{Correlation1} \\
\Gamma_{2}(t) &=&\int_{0}^{t}d\tau \int_{0}^{\infty }d\omega
J\left( \omega \right) e^{-i\left( \omega -\omega _{L}+\delta \right) \left(t-\tau \right) },
\label{Correlation2} \\
\Gamma_{3}(t) &=&\int_{0}^{t}d\tau \int_{0}^{\infty }d\omega
J\left( \omega \right) e^{-i\left( \omega -\omega _{L}-\delta \right) \left(t-\tau \right) }.
\label{Correlation3}
\end{eqnarray}
The master equation (\ref{Equation9}) is a convolutionless time-local differential equation. The time-dependent coefficients
$\Gamma_{1}(t)$, $\Gamma_{2}(t)$, and $\Gamma_{3}(t)$ account for the memory effect of the non-Markovian environment.
To characterize the environment and calculate these time-dependent coefficients in equations (\ref{Correlation1})-(\ref{Correlation3}),
we must consider a spectral density to characterize the structured environment. The time-dependent correlation functions fully characterize
the non-Markovian memory effect given the spectral density $J(\omega)$. Here we consider an Ohmic spectral density
\cite{leggett1987dynamics,haikka2013non,addis2014coherence,chen2014investigating,chen2016enhanced}, which is a
common class of spectral density that simulates the dynamics of a two-level open quantum system
\begin{equation}
J\left( \omega \right) =\gamma \omega \exp \left( -\omega /\lambda \right).
\end{equation}
The time-dependent coefficients (\ref{Correlation1})--(\ref{Correlation3}) appearing in the master equation contains the
non-Markovian characteristics of the open quantum system. Here $\gamma$ is the coupling strength between the system and
dissipative bath and is measured in units of $\gamma_{0}$ which is a fixed frequency, closely related to the relaxation time of the
qubit. The constant $\gamma_0$ is the relaxation rate related to the $T_1$ relaxation time of the qubit under spontaneous decay.
Considering a typical experimental value of the relaxation time of a superconducting qubit
\cite{burnett2019decoherence,krantz2019quantum,etxezarreta2021time}, we find the relaxation rate
$\gamma_0$ to be of the order of MHz. In this work, we have taken $\gamma=0.1\gamma_{0}$
for all results where coupling strength $\gamma$ is fixed. The parameter $\lambda$ is the cutoff frequency of the bath spectrum.
We see two distinct types of qubit dynamics based on the value of the system-environment parameters. For $\lambda > 2 \gamma$,
the reservoir correlation time is very short compared to the relaxation time of the qubit and consequently the dynamics is Markovian.
When $\lambda < 2 \gamma$, the reservoir correlation time is comparable to the relaxation time of the qubit and hence we observe non-Markovian effects. For the Markov regime, we consider a value of $\lambda=5\gamma_{0}$, and in the non-Markov regime
$\lambda=0.01\gamma_{0}$. We investigate the phase synchronization of the qubit for two different values of the detuning namely
$\Delta=0$ and $\Delta=\gamma_0$.

\section{Husimi $Q$-function}\label{sec:Husimi}

\begin{figure}[h]
\includegraphics[width=\linewidth]{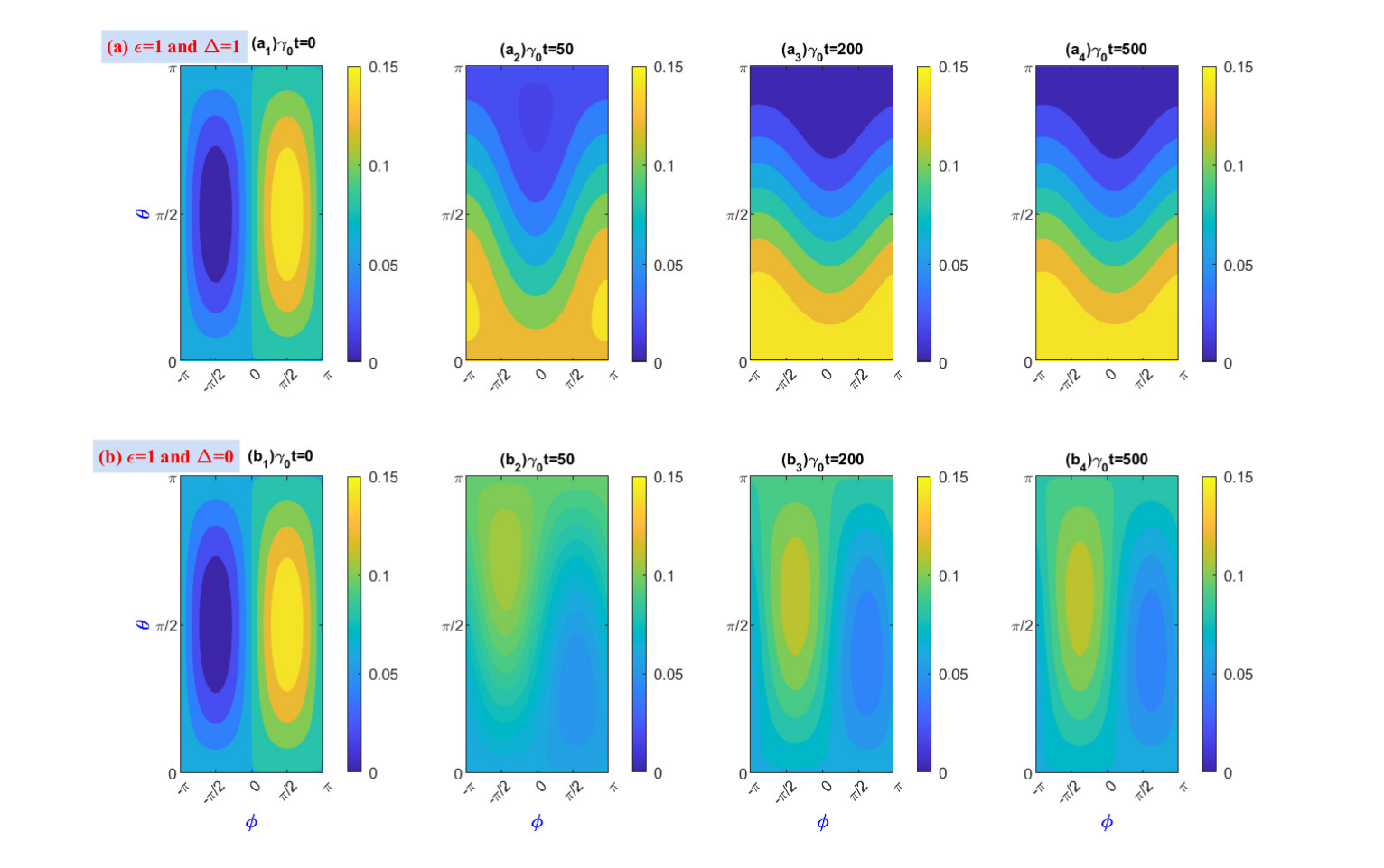}
\caption{Temporal evolution of the Husimi $Q$-function of a single qubit coupled to a Makovian dissipative
bath and under an external driving is shown in the plot for (a) $\epsilon =1$ and $\Delta =1$ (b) $\epsilon =1$ and
$\Delta =0$. We have taken the spectral width $\lambda=5\protect\gamma _{0}$ with four different evolution times.
The system-environment coupling strength is fixed at $\gamma=0.1\gamma _{0}$.}
\end{figure}
\begin{figure}[h]
\includegraphics[width=\linewidth]{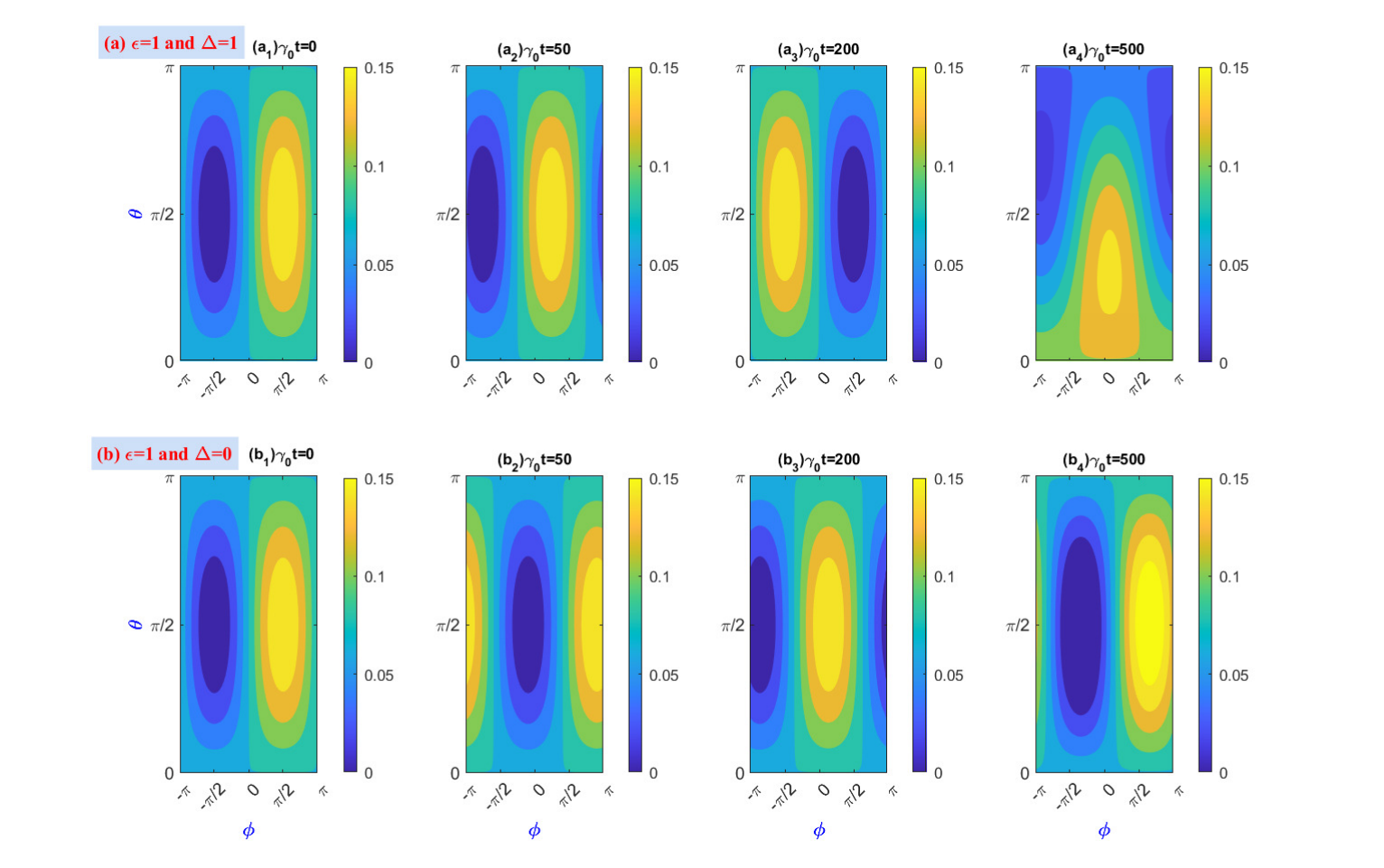}
\caption{Temporal evolution of the Husimi $Q$-function of a single qubit coupled to a non-Makovian dissipative
bath and under an external driving is shown in the plot for (a) $\epsilon=1$ and $\Delta=1$ (b)
$\epsilon=1$ and $\Delta=0$. We have taken the spectral width $\lambda=0.01 \gamma_{0}$
with four different evolution times. The system-environment coupling strength is taken as $\gamma=0.1\gamma _{0}$.}
\end{figure}
The phase synchronization of the driven TLS can be characterized using the Husimi $Q$-function \cite{husimi1940some,gilmore1975classical}
which is a quasi-probability distribution capable of capturing the phase space dynamics.  In our work we use the spin-coherent states
and the corresponding expression reads:
\begin{equation}
Q\left( \theta,\phi,t\right) = \frac{1}{2\pi}\left\langle \theta,\phi \right\vert \rho \left(t\right) \left\vert \theta,\phi \right\rangle,
\label{Qfunction}
\end{equation}
where $\rho(t)$ is the time evolved density matrix.  For the two level system $\vert \theta,\phi \rangle$ are the spin-coherent states
which are eigenstates of the spin operator $\sigma _{n}=\mathbf{\hat{n}}\cdot \hat{\sigma}$ along the unit vector
$\mathbf{\hat{n}}$ with polar coordinates $\theta$ and $\phi$.  This state $\left\vert \theta,\phi \right\rangle$
$=\cos\left( \theta /2\right) \left\vert 1\right\rangle +\sin \left( \theta /2\right) e^{i\phi }\left\vert 0\right\rangle$
represents a point on the surface of a Bloch sphere, where $|0 \rangle$ and $|1 \rangle$ are the eigenstates of the spin operator $\sigma_{z}$.
Once the temporal evolution of reduced density matrix are determined from Eq.~(\ref{Equation9}), it is easy to obtain the time dynamics
of $Q$-distribution as a function $\theta$, $\phi$ and $t$ as follows:
\begin{equation}
Q\left( \theta ,\phi ,t\right) = \frac{1}{2\pi }\left[
\cos ^{2}\left( \theta /2\right) \rho _{11}\left( t\right) + \cos \left(\theta /2\right)
\sin \left( \theta /2\right) e^{i\phi }\rho _{10}\left(t\right)
+\cos \left( \theta /2\right) \sin \left( \theta /2\right) e^{-i\phi }\rho
_{01}\left( t\right) +\sin ^{2}\left( \theta /2\right) \rho _{00}\left(t\right) \right].
\label{Eq11}
\end{equation}
From Eq.~(\ref{Eq11}) we can see that the $Q$-function can be expressed as a weighted sum of the individual density matrix
elements.

An analysis of the transient dynamics of $Q\left( \theta,\phi,t\right)$ for the Markovian and non-Markovian evolution is presented
in Fig.~1 and Fig.~2 respectively for an initial state of $\left\vert +\right\rangle=\left( \left\vert
0\right\rangle +\left\vert 1\right\rangle \right) /\sqrt{2}$. In Fig.~1, the Husimi $Q$-function for the Markov regime
$(\lambda=5\gamma_{0})$ is shown for two values of the detuning parameter namely (a) $\Delta=1$ and (b) $\Delta=0$
corresponding to a finite detuning and zero detuning respectively. The driving field strength ($\epsilon$) is set to be one
in the computations. The series of plots in Fig.~$1(a)$ from $(a_{1})$ to $(a_{4})$ trace the evolution of $Q$-function for
$\Delta=1$. At time $t=0$, the $Q$-function is nonuniformly distributed and it is peaked at $\phi=0$. This non-uniform
phase distribution implies the presence of a phase preference in the system. As the system evolves in time, the
phase distribution becomes more and more uniform and in the long-time limit the phase preference is completely wiped out.
The set of plots Fig.~$1(b_{1})$ to $1(b_{4})$ projects the phase distribution dynamics for a TLS with zero detuning.
Here we find that the phase preference survives over a longer time and for $\gamma_{0} t=500$ the phase localization peak
i.e., the maximal value of the phase localization shifts towards $\phi = - \pi/2$.

The phase preference of the driven TLS in the non-Markovian regime ($\lambda=0.01\gamma_{0}$) has also been shown for
(a) finite detuning ($\Delta=1$) and (b) zero detuning ($\Delta=0$) and the results are shown in Fig.~$2(a)$ and
Fig.~$2(b)$ respectively. The series of plots from Fig.~$2(a_{1})$ to  $2(a_{4})$ show the evolution of the localized peak
of the Husimi $Q$-function for a finite detuning. Here we find that compared to Markov case, the phase delocalization
process is significantly slowed down for non-Markov dynamics. The series of plots in Fig.~$2 b_{1}$ to $2 (b_{4})$ show the evolution
of the $Q$-function for zero detuning. While the relative position of the peak shifts in time, the phase does not spread out.
Hence we can conclude that the phase localization is present in the Markovian and non-Markovian regime for zero detuning.

The relative phase in a quantum system is encoded in the off-diagonal elements.  Hence any phase localization will imply that the
off-diagonal elements remain constant in their value.  To verify this we plot the evolution of $|\rho_{10}(t)|$ for both Markovian
and non-Markovian evolution in Fig. (3).  For the Markovian evolution described in Fig 3(a), the off-diagonal elements oscillate
in the short time limit and saturates to a finite value in the long time limit.  The saturation value for the condition
$\Delta =0$ is comparatively higher than when there is a finite detuning ($\Delta = 1$), but is lower than that of the
initial value.  Hence we conclude from Fig 3(a) that the phase localization is significantly degraded due to the Markovian nature
of the bath.  In the case of non-Markovian evolution, the value of the off-diagonal element remains almost constant throughout
the evolution and is the same for both zero detuning ($\Delta = 0$) and finite detuning ($\Delta = 1$) situations.  Thus the
driven TLS exhibits phase localization in the non-Markovian regime due to information backflow from the external environment.

It is well-known that the off-diagonal elements signal the presence of quantum coherence in the system.  Hence we conclude
that the quantum phase localization and consequently the quantum phase synchronization is due to long lasting quantum coherence
in the system. This relationship between quantum phase synchronization and quantum coherence has been examined in detail in
Ref.~\cite{ali2024detecting}. The synchronization in our work is quite different from the ones discussed in
Ref.~\cite{karpat2019quantum,karpat2021synchronization}.  In these references, there are two coupled quantum systems of which one
is connected to an external bath and we study the synchronization of the systems which is not in contact with the bath and
this type of synchronization is referred to as spontaneous mutual synchronization.

\begin{figure}[h]
\includegraphics[width=\linewidth]{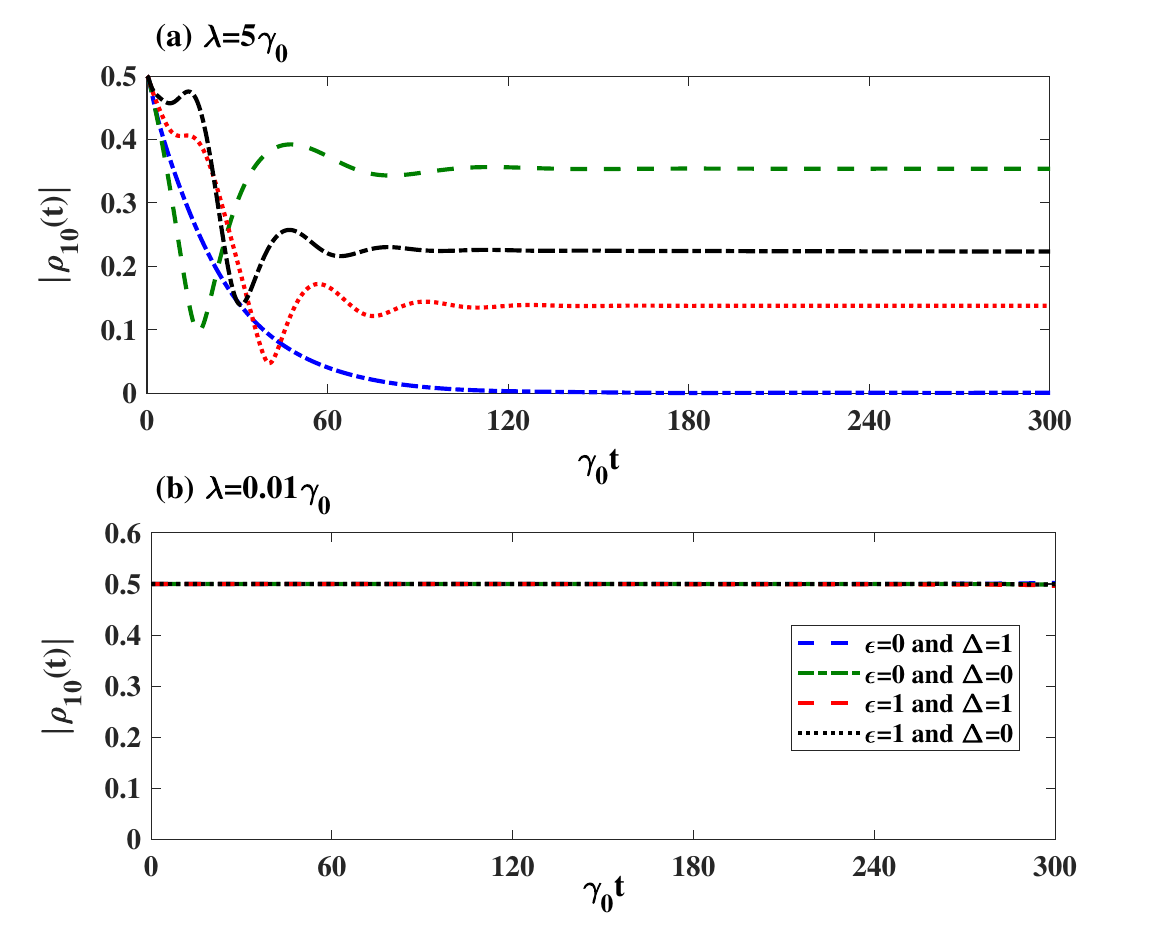}
\caption{The plot of the off-diagonal element $\left\vert \rho_{10}\left( t\right) \right\vert$
as a function of time is given for (a) Markov and (b) non-Markovian evolution.
The system-environment coupling strength is taken as $\gamma=0.1\gamma _{0}$.}
\end{figure}

\begin{figure}[h]
\includegraphics[width=\linewidth]{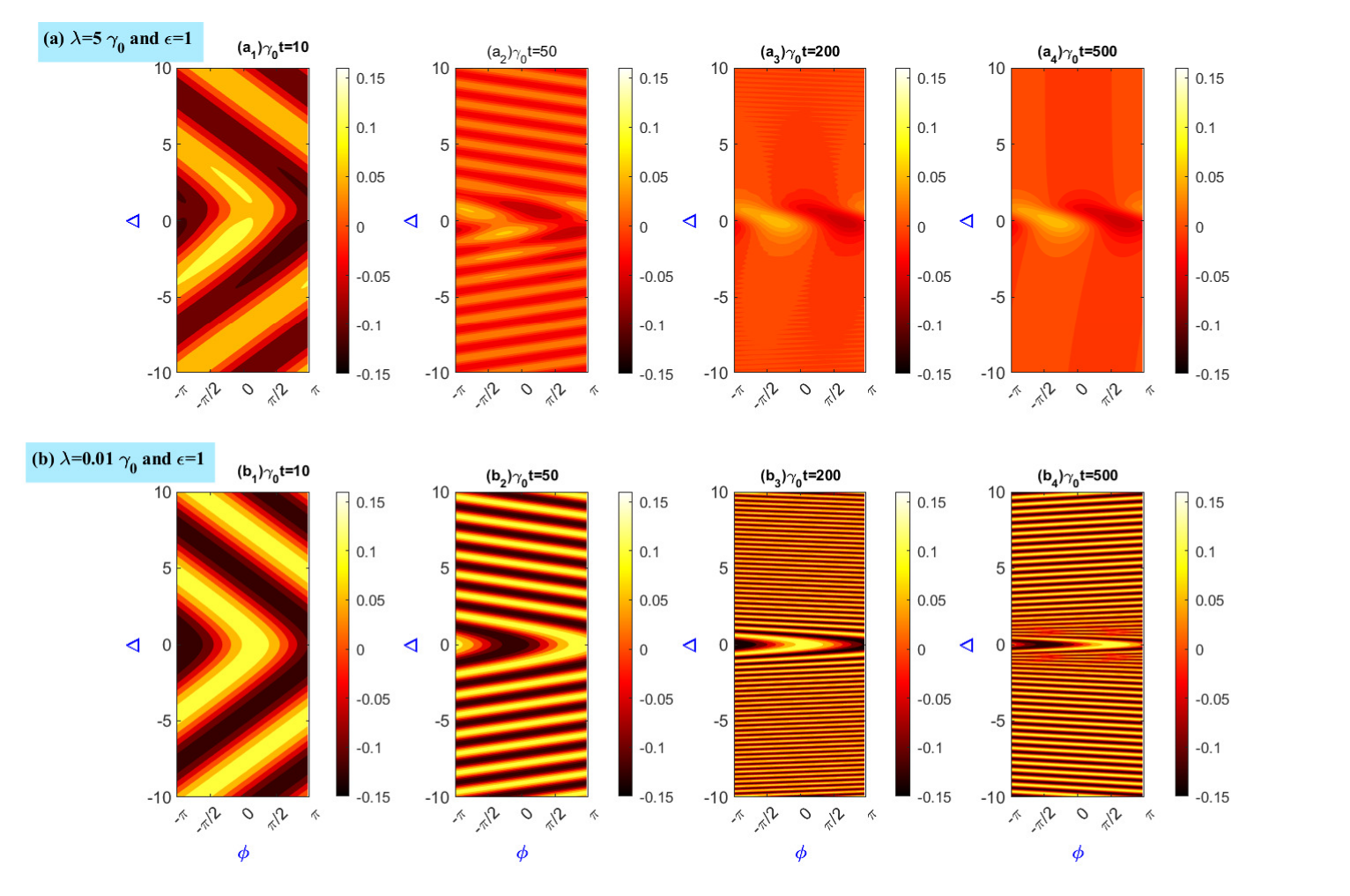}
\caption{The contour plot of the phase shift distribution $S\left(\phi,t\right)$ as a function of
$\Delta$ and $\phi$ for different evolution times $\gamma_{0}t$ is shown with (a)
$\epsilon=1$ in Markov case ($\lambda=5\gamma_{0}$) and (b) $\epsilon=1$ in
non-Markovian case ($\lambda=0.01\gamma_{0}$). The system-environment coupling strength is taken as $\gamma=0.1\gamma_{0}$.}
\end{figure}

\section{Synchronization measure and Arnold's tongue}\label{sec:arnold}
\subsection{Shifted phase distribution}
The Husimi $Q$-function can be used to identify the presence of phase localization in the system. To find the strength of
the phase distribution measure $P(\phi, \rho)$ proposed in Ref.~\cite{roulet2018synchronizing,koppenhofer2019optimal,parra2020synchronization}. This function is obtained by
integrating the $Q$-function over the angular variable $`\theta'$ for a given $\rho$. A quantum state with uniform phase
distribution has $P(\phi, \rho_{0})=-1/2 \pi$.  So the non-uniformity in hte phase distribution or rather
the phase preference can be measured using a function called the shifted phase distribution given below:
\begin{equation}
S\left(\phi,t\right) \equiv P(\phi, \rho) -  P(\phi, \rho_{0})
= \int_{0}^{\pi }Q\left(\theta,\phi,t\right) \sin \theta d\theta - \frac{1}{2\pi }.
\label{Sfunction}
\end{equation}
This function is zero iff there is no phase localization which implies there will not be any phase synchronization.
Evaluating the integral over the angular variable $\theta$ and using the condition of trace invariance
i.e., $\rho_{00}(t) + \rho_{11}(t) = 1$ we get:
\begin{equation}
S\left(\phi,t\right) =\frac{1}{8}\left[ \rho _{10}\left( t\right) e^{i\phi} + \rho _{01}\left( t\right) e^{-i\phi }\right].
\label{Sfunction2}
\end{equation}
We plot the shifted phase distribution $S(\phi,t)$ as a function of $\Delta$ and $\phi$ at different instants of time
in both the Markovian and non-Markovian regime.

In Fig. $4(a)$ we plot the Markovian dynamics $(\lambda=5 \gamma_{0})$
in which the subplots Fig.~$4 (a_{1})$ to  Fig.~$4 (a_{4})$ describe the variation of $S(\phi,t)$ with $\Delta$ and
$\phi$ for the time intervals of $\gamma_{0}t=10, 50, 200, 500$ respectively. From the plot Fig.~$4 (a_{1})$ we find
that there are regions of phase localization for $\gamma_{0}t=10$.  With increase in $\gamma_{0}t$, we can observe that the
regions where value of $S(\phi,t)=0$ increases and for $\gamma_{0}t=500$ the shifted phase is distribution is uniformly
zero everywhere.  The shifted phase distribution of non-Markovian dynamics $(\lambda=0.01 \gamma_{0})$ is plotted in
Fig. $4 (b)$. The subplots from Fig. $4 (b_{1})$ to  Fig. $4 (b_{4})$ show $S(\phi,t)$ as a function of $\Delta$ and
$\phi$ for the time intervals of $\gamma_{0}t=10, 50, 200, 500$ respectively.  Here again at $\gamma_{0}t=10$ show
in Fig $4 (b_{1})$ we find that there are regions of phase localization. While the characteristics of this region
change with time, the shifted phase distribution never becomes zero uniformly for all values of $\Delta$ and $\phi$
even when $\gamma_{0}t=500$.  There are periodic finite values of detuning where the phase synchronization occurs.
Hence our results show that the phase localization disappears in the long time limit for Markovian dynamics.
In the case of non-Markovian dynamics the phase localization is present even in the long tme limit.

\subsection{Arnold tongue}
Another measure that is widely used to characterize quantum synchronization is the maximum of the shifted phase distribution.
This value can be found by computing $S(\phi,t)$ over the entire range of $\phi$ and finding the maximal value.  For the
single qubit system (TLS) which we are investigating the expression for the maximal shifted phase distribution $S_{m}(t)$ reads:
\begin{equation}
  {\rm max}~S(\phi,t) \equiv S_{m} (t) =  \frac{1}{4} \vert \rho_{10} \vert  = \frac{1}{8} C_{\ell_{1}}(\rho)
  \label{maximumshiftedphasedistribution}
\end{equation}
where $C_{\ell_{1}}(\rho)$ is the $\ell_{1}$-norm measure of quantum coherence. This connection between quantum phase localization
and quantum coherence was discussed in detail in Ref.~\cite{ali2024detecting}. This is deemed to be a natural connection because
the relative phases which are being localized are the same ones which describe the quantum coherence in the system.

\begin{figure}[h]
\includegraphics[width=\linewidth]{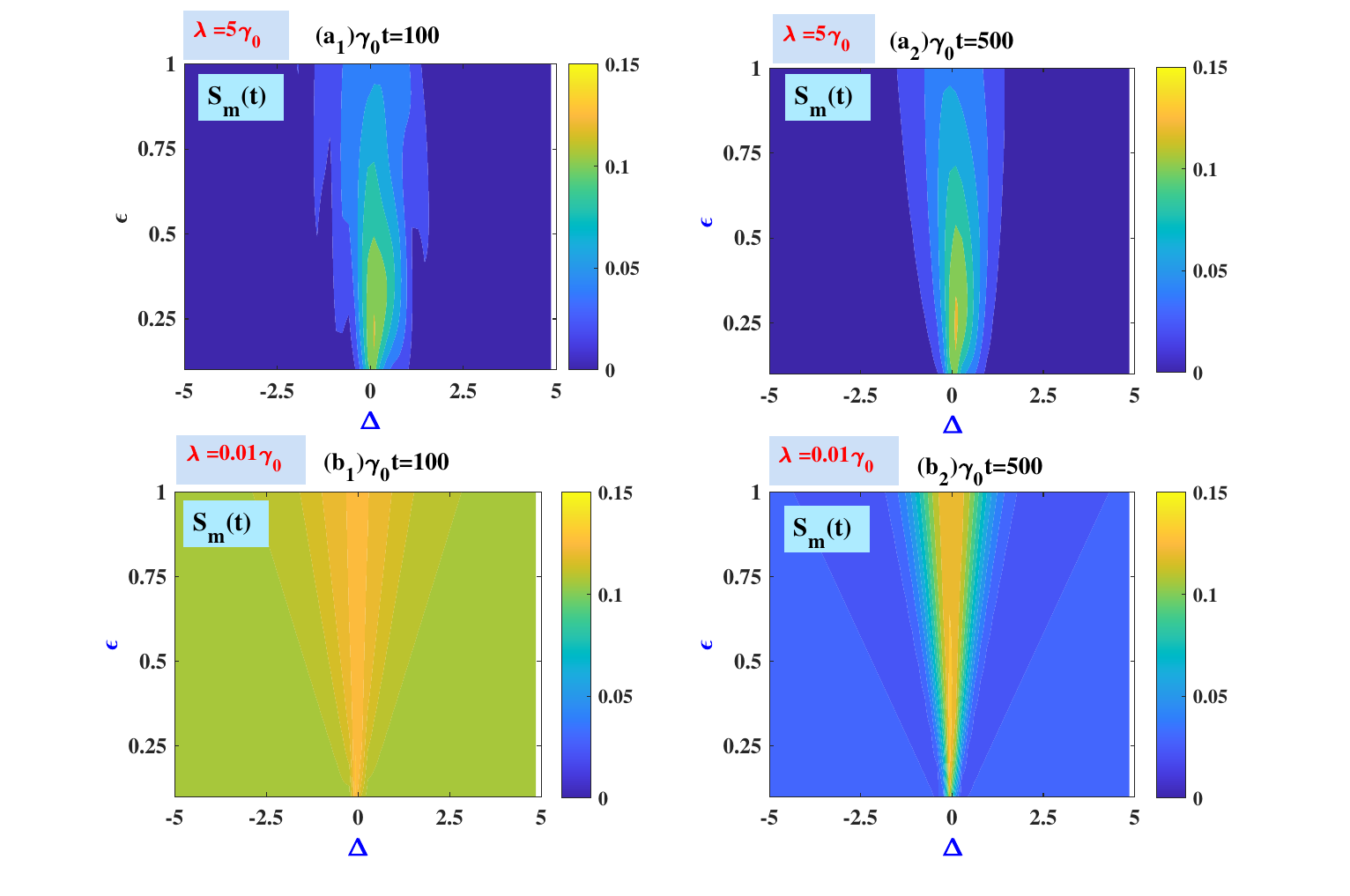}
\caption{Contour plot of the maximal value of the shifted phase distribution $S_{m}\left(t\right)$
as a function of $\Delta$ and $\epsilon$ in the Markovian regime ($\lambda=5\gamma_{0}$)
for two evolution times (a$_{1}$) $\gamma_{0}t=100$ and (a$_{2}$) $\gamma_{0}t=500$.
On the other hand, $S_{m}\left(t\right)$ as a function of $\Delta$ and $\epsilon$ is shown
in the non-Markovian regime ($\lambda=0.01\gamma_{0}$) with varying evolution times
(b$_{1}$) $\gamma_{0}t=100$ and (b$_{2}$) $\gamma_{0}t=500$. The system-environment
coupling strength is taken as $\gamma=0.1\gamma_{0}$.}
\end{figure}
\begin{figure}[h]
\includegraphics[width=\linewidth]{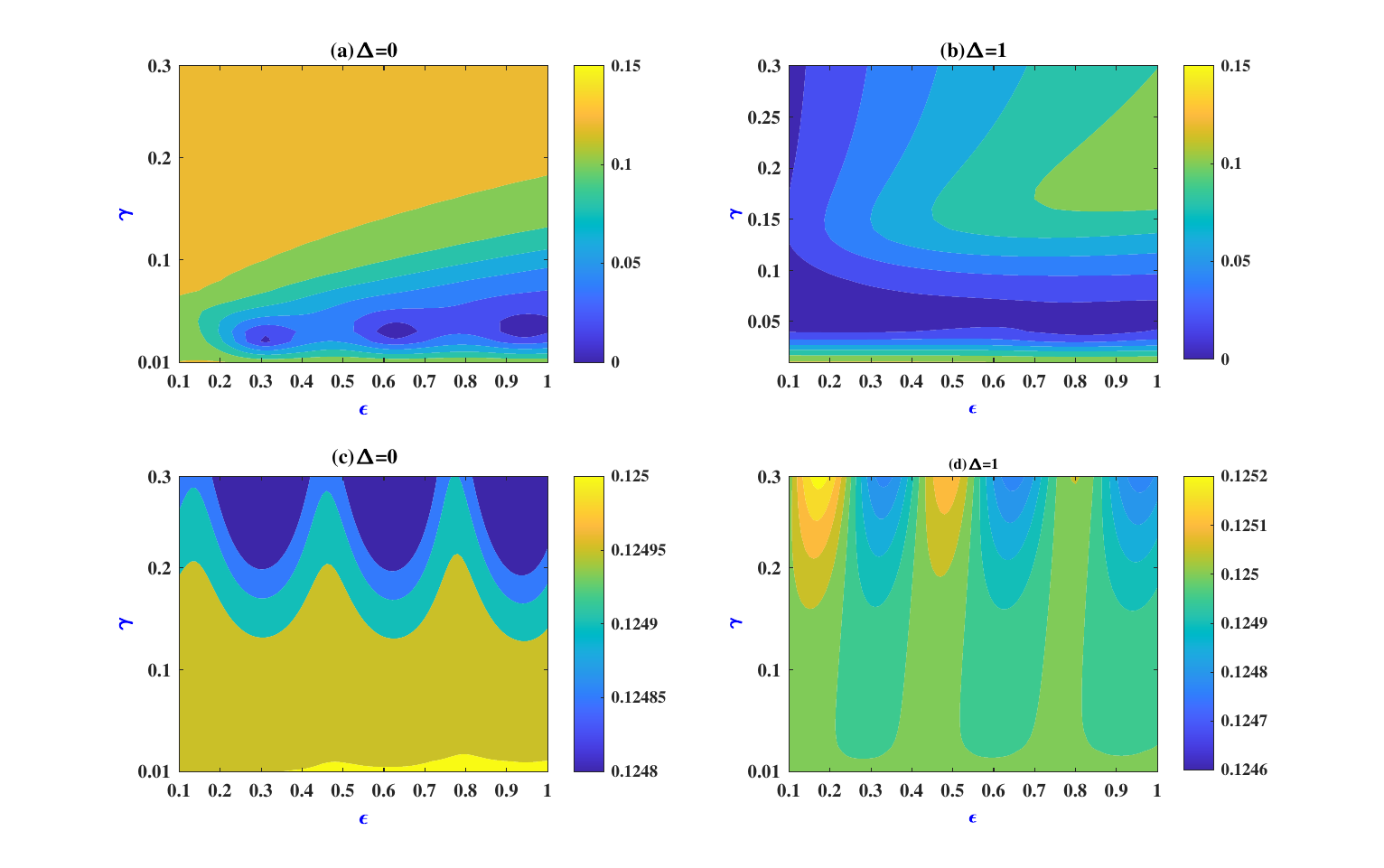}
\caption{3D contour plot of $S_{m}(t)$ as a function of $\epsilon$ and $\gamma$ is shown in the
Markovian regime ($\lambda=5\gamma _{0}$) for (a) $\Delta=0$ and (b) $\Delta=1$.
Next, $S_{m}\left(t\right)$ as a function of $\epsilon$ and $\gamma$ is shown in the non-Markovian regime
($\lambda=0.01\gamma_{0}$) for (c) $\Delta =0$ and (d) $\Delta=1$. The time evolution is fixed at
$\gamma_{0}t=500$.}
\end{figure}

A contour plot of $S_{m}(t)$ as a function of the detuning parameter $\Delta$ and laser driving strength $\epsilon$ is
shown in Fig.~5 for both the Markovian as well as the non-Markovian regime.  For the Markovian regime $(\lambda=5 \gamma_{0})$
are shown through the plot in Fig.~5 (a) where $S_{m}(t)$ is given for $\gamma_{0} t=100$ in Fig.~5 $(a_{1})$ and for
$\gamma_{0} t=500$ in Fig.~5 $(a_{2})$ respectively.  From the plots we observe certain regions where $S_{m}(t)$ is positive,
indicating that there is a region where the phases are locked, i.e., they remain invariant.  The investigations corresponding
to the non-Markovian regime $(\lambda=0.01 \gamma_{0})$ is shown through the set of plots in Fig.~5 (b) where Fig.~5 $(b_{1})$
gives the features for $\gamma_{0} t=100$ and Fig.~5 $(b_{2})$ shows the result for $\gamma_{0} t=500$ respectively.  In
Fig.~5 $(b_{1})$ we find that there is a phase locking present for the whole region.  But when we get to Fig.~5 $(b_{2})$ we
notice that the phase locking is zero for most values of $\Delta$. Also we observe a triangular region where the phase is
locked, a feature which is generally referred to as Arnold tongue and is considered as a signature of quantum phase synchronization.

To get a complete picture of phase locking we plot $S_{m}\left(t\right)$ as a function of the strength of the laser driving
($\epsilon $) and the system-bath coupling strength ($\gamma$) as well.  The results are obtained for both the Markovian
regime $(\lambda=5 \gamma_{0})$ as well as for the non-Markovian regime $(\lambda=0.01 \gamma_{0})$ in Fig. 6 respectively.
From the plots for both Markovian and non-Markovian regimes we find regions where the phase is locked. This reinforces the
conclusion obtained through Fig.~5.

\subsection{Limit cycle analysis}
The discussions we had so far show that the phase is localized in the driven two-level system which we are currently
investigating.  In our earlier work in Ref.~\cite{ali2024detecting} we have shown that while phases can be localized it
does not always lead to phase synchronization. In fact we show that the presence of Arnold tongue does not imply that
there is a phase synchronization in the system. This necessitates us to find a new way to observe and analyze quantum
phase synchronization.

An alternative method to confirm the presence of quantum phase synchronization is through the existence of a limit cycle.
In the long-time limit if the trajectory of the driven TLS system becomes a closed trajectory, then the system is said
to have a limit cycle. In our work, we have considered a single qubit and investigated it in the spin coherent basis.
Here the trajectory of the qubit in the Bloch sphere can capture its dynamics in the phase space. Below we present a
study of the dynamics of the TLS on a three-dimensional sphere and discuss our results.

\begin{figure}[h]
\includegraphics[width=\columnwidth]{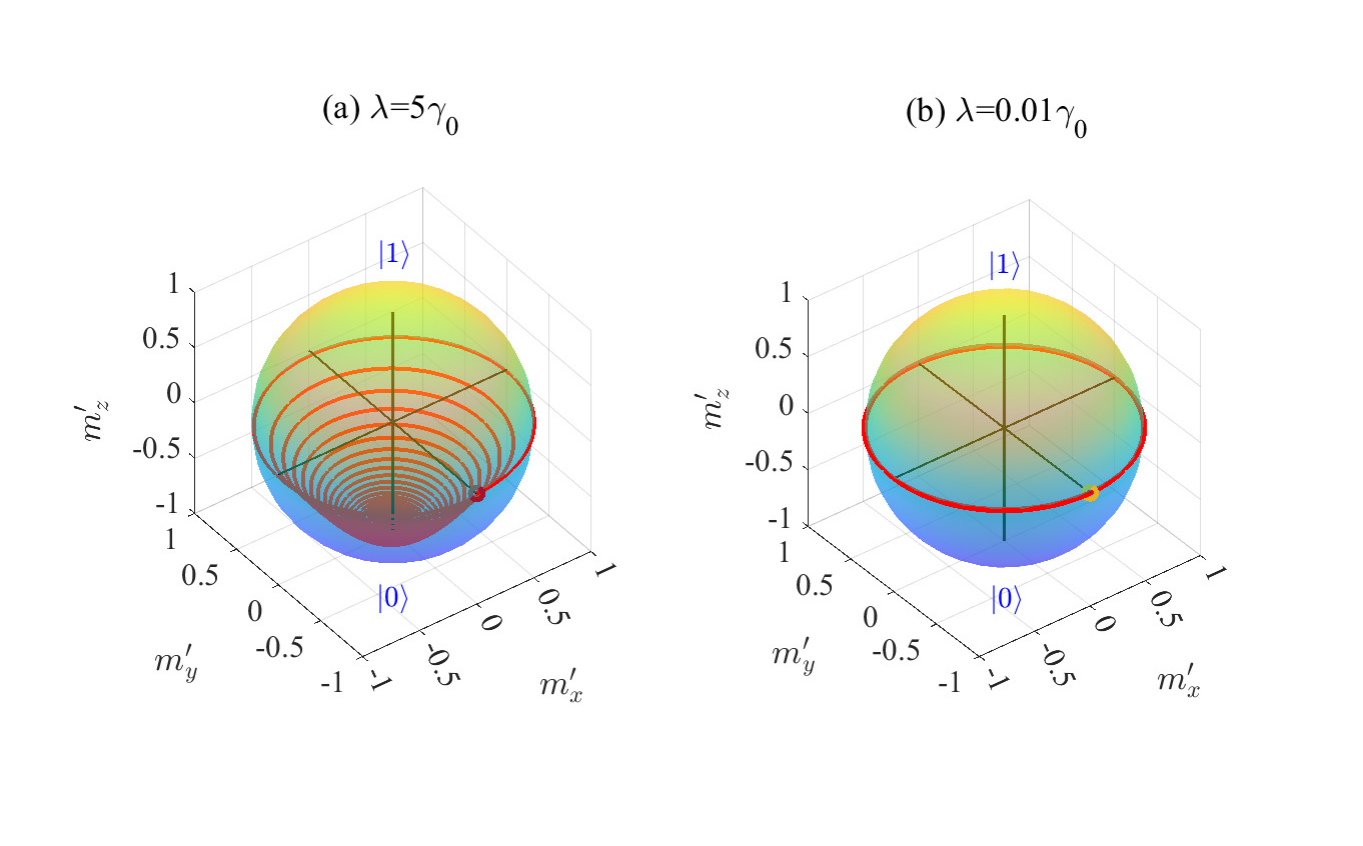}
\caption{Qubit trajectories on the Bloch sphere is shown for Markovian dynamics in
(a) with $\lambda=5\gamma_0$ and for non-Markovian dynamics in
(b) with $\lambda=0.01\gamma_0$. Values of the other parameters are taken as
$\gamma=0.1\gamma _{0}$, $\epsilon=\gamma_0$, and $\Delta=\gamma_0$.}
\label{fig7}
\end{figure}

To plot the qubit trajectories on the Bloch sphere we need to consider the system in the nonrotating frame of reference.
For this we transform the qubit density matrix with the unitary transformation $U=e^{\frac{i}{2}{\sigma }_{z}\omega_{L}t}$ and
perform the transformation $\rho^{\prime}=U^{\dagger} \rho U$ where $\rho^{\prime}$ is the density matrix in the
nonrotating frame of reference. The Bloch vector components $(m_{x}^{\prime}, m_{y}^{\prime}, m_{z}^{\prime})$ in the nonrotating
frame can be expressed as follows:
\begin{subequations}
\begin{align}
\label{mx}
& m_x^{\prime} = m_x \cos(\omega_L t) - m_y \sin(\omega_L t), \\
\label{my}
& m_y^{\prime} = m_x \sin(\omega_L t) + m_y \cos(\omega_L t), \\
\label{mz}
& m_z^{\prime} = m_z,
\end{align}
\end{subequations}
where $m_x={\rm Tr}(\sigma_x \rho)$, $m_y={\rm Tr}(\sigma_y \rho)$, $m_z={\rm Tr}(\sigma_z \rho)$ are the Bloch
vector components in the rotated frame. To visualize the trajectory we observe the dynamics of the two-level system and
for this we plot the Bloch vector components (\ref{mx}-\ref{mz}) of the time evolved reduced density matrix of the
qubit up to $\gamma_0 t=300$ for both Markovian and non-Markovian evolution in Fig.~(\ref{fig7}). The Markovian
dynamics is captured in Fig.~\ref{fig7}(a) for $\lambda=5\gamma_0$. We observe that the initial state
$|+\rangle = (|0 \rangle + |1 \rangle) / \sqrt{2}$ evolves to a $|0\rangle$ state. The trajectory is not closed and the final
steady state is a point on the sphere. On the contrary the trajectory of the non-Markovian dynamics is captured in Fig.~\ref{fig7}(b)
for $\lambda=0.01\gamma_0$, which shows that the dynamics evolves such that the trajectory is a closed curve. This feature
indicates that the quantum phase of the two level system is synchronized. Thus the limit cycle is established
for the two-level system under driving and in the long-time limit the system precesses around the $z$-axis.

\section{Conclusion}\label{sec:conclusion}

In this work, we have investigated quantum phase synchronization for a two-level system (qubit) driven by a
semiclassical laser field in the presence of both Markovian and non-Markovian dissipative environments. We used
the Husimi Q-function to show that quantum phase synchronization is significantly enhanced in the non-Markovian regime.
In the Markov regime with finite $\Delta$, as the system evolves in time, the phase distribution becomes more and more
uniform and in the long-time limit, the phase preference is completely wiped out. For $\Delta=0$, the system can be
synchronized even in the Markovian regime in presence of the driving field. For finite detuning ($\Delta=1$), we find
that the phase preference survives over a longer period in the non-Markovian dynamics. To quantify the phase
synchronization, we plot the shifted phase distribution and its maximum value for a wide range of system-environment
parameters. The maximum of the shifted phase distribution exhibits an Arnold tongue like feature, demarcating the phase
localized and phase delocalized regions in the system environment parameter space. We establish a mathematical connection
between the maximum of the shifted phase distribution and quantum coherence given by Eq.~(\ref{maximumshiftedphasedistribution}).
The trajectory of the qubit in the Bloch sphere captures its dynamics in the phase space. In the long-time limit the trajectory
of the driven system becomes a closed trajectory, hence the system is said to have a limit cycle. We confirm the presence of
quantum synchronization through the existence of a limit cycle. Using a shifted phase distribution $S(\phi,t)$, we show that
quantum phase synchronization is significantly enhanced in the non-Markovian regime. To quantify the phase synchronization,
we plot the shifted phase distribution and its maximum value for a wide range of system-environment parameters. We plot the
maximum of the shifted phase distribution in two different ways: (a) as a function of the detuning parameter ($\Delta$) and laser
driving strength ($\epsilon$) and (b) as a function of the strength of the laser driving ($\epsilon$) and the system-bath coupling
strength ($\gamma$). We systematically discussed how the synchronization regions are determined by various system-environment
parameters and observed the typical Arnold tongue features of a phase synchronized qubit. The driven qubit is synchronized inside
the tongue region and desynchronized outside the Arnold tongue.

\begin{acknowledgments}
PWC received funding from the Department of Physics, National Atomic Research Institute (NARI), Taiwan.
\end{acknowledgments}

\vskip 0.5cm

{\bf Data availability statement}: All data that support the findings of this study are included within the article
(and any supplementary files).

\vskip 0.5cm

{\bf Author Contributions}: All the authors contributed equally to the concept, calculation, writing, and interpretation
of the present work.

\bibliography{SyncDriven.bib}

\end{document}